\documentclass[letterpaper]{article}
\usepackage{aaai2026}
\nocopyright
\usepackage{times}  % DO NOT CHANGE THIS
\usepackage{helvet}  % DO NOT CHANGE THIS
\usepackage{courier}  % DO NOT CHANGE THIS
\usepackage[hyphens]{url}  % DO NOT CHANGE THIS
\usepackage{graphicx} % DO NOT CHANGE THIS
\usepackage{natbib}  % DO NOT CHANGE THIS AND DO NOT ADD ANY OPTIONS TO IT
\usepackage{caption} % DO NOT CHANGE THIS AND DO NOT ADD ANY OPTIONS TO IT
\usepackage{amsmath,amssymb}   % math in the measures and results
\usepackage{booktabs}          % \toprule, \midrule, \bottomrule in every table
\usepackage{xcolor}            % \color{teal} in the Paper Checklist answers only (no colour in the text)

\usepackage{nameref}

\title{Attentional DoS: Repeat Reposting, Collective Attention, and Information Diffusion on X}
\author{
    Genta Toya\textsuperscript{\rm 1},
    Bruno T. Sugano\textsuperscript{\rm 1},
    Kei Ichikawa\textsuperscript{\rm 1},
    Qianyun Wu\textsuperscript{\rm 2,\rm 1},
    Shuhei Saigusa\textsuperscript{\rm 1},\\
    Yasuhiro Hashimoto\textsuperscript{\rm 3},
    Masashi Toyoda\textsuperscript{\rm 4},
    Naoki Yoshinaga\textsuperscript{\rm 4},
    Kazutoshi Sasahara\textsuperscript{\rm 1}
}
\affiliations{
    \textsuperscript{\rm 1}Institute of Science Tokyo, Tokyo, Japan\\
    \textsuperscript{\rm 2}Hirosaki University, Hirosaki, Japan\\
    \textsuperscript{\rm 3}University of Aizu, Aizuwakamatsu, Japan\\
    \textsuperscript{\rm 4}Institute of Industrial Science, The University of Tokyo, Tokyo, Japan
}

\newcommand{\ACKTEXT}{This paper is based on results obtained from a project, JPNP22007, commissioned by the New Energy and Industrial Technology Development Organization (NEDO).}
\newcommand{\CHECKLIST}{}

\begin{document}
\maketitle
\begin{abstract}
Collective attention is a finite shared resource, and social media posts compete for limited opportunities to be seen. On X, users can undo a repost and repost it again. By repeating this cycle, the same user can put the same post back into the followers' timelines any number of times without making new content. We call this procedure repeat reposting, and we read it as placing repeated demand on this shared resource (Attentional DoS). We formalize this idea and conduct an exploratory analysis of repeat reposting in a large-scale dataset of cascades with at least 1,000 reactions, originating from posts classified as Japanese on X, covering April 2025 to March 2026. Repeat reposts are rare: they appear in only a small share of all (user, post) pairs. Even so, close to a million posts have at least one repeater, and most repeats come from a small group of habitual accounts. The central result is that subsequent audience growth is associated less with the number of repeats than with the estimated reach of the repeating accounts. When we look at all repeat reposts through seriality, how habitually the same groups of accounts repeat reposts across many posts, a distinct distributed form emerges: several serial amplifiers converge on the same post (Attentional DDoS). Its synchrony, how closely their actions are timed together, forms a continuum from bursts on the scale of minutes to a daily clock. A check of the content of 99.7\% of amplified posts shows that most ADDoS repeat events are directed at Chinese-script content, which our vocabulary matching and sample inspection indicate is predominantly commercial spam. Repeat reposting enables repeated re-entry into the competition for visibility without creating new content. The observed pattern is better characterized as repeated temporal coverage of an existing audience and the self-reinforcement of posts that are already growing, rather than as evidence that repeat reposting takes reach from other content.
\end{abstract}

% ---------- 01-intro ----------
\section{Introduction}\label{sec:intro}

\subsection{Scarce Collective Attention and the Competition for Visibility}\label{sec:measurement}
In a world rich in information, attention becomes the scarce resource, because information consumes it \cite{simon1971}. A social media timeline gives a large number of posts a limited number of chances to be seen, and posts compete for those chances (see \nameref{sec:related}). This paper looks at an operation through which the same post can be reposted repeatedly over time: repeat reposting, which puts the same post back into the followers' timelines any number of times.

\subsection{Attentional DoS: Repeated Requests on a Finite Shared Resource}\label{sec:bg}
On X, users can undo a repost and repost it again. By repeating this cycle of repost, undo, and repost again, the same user can put the same post back into the followers' timelines any number of times; each creates another repost event for the same post as a new share rather than as new content (Figure~\ref{fig:schematic}, top). Whether this repetition really brings new readers (in our data, new unique reactors) is an empirical question, separate from whether the mechanism exists.

Repeat reposting can be seen as placing repeated demand on a finite shared resource, collective attention. We call this Attentional DoS, by structural analogy with denial-of-service attacks on computer networks. The point of the analogy is the structure: an excess of requests on a finite shared resource. A classical DoS attack exhausts a server's finite resources with requests. Informational DoS (IDoS) \cite{huang2021,huang2022} is defined as an attack in which the attacker generates a large volume of feints (low-cost decoy actions, real in themselves), floods security operators with alerts, wastes their cognitive resources, and so keeps them from finding the real attacks hidden among the feints. Attention is already the resource being exhausted there, and this paper follows that position and extends it from the security operator to the public information ecosystem: attention is a finite shared resource, rivalrous in use, and this is what motivates the word ``denial''. The analogy has a limit, however: in IDoS the feints hide a real attack, while repeat reposting hides nothing behind the repeats, so the repetition is the whole of the operation. What we actually measure is the existence, scale, and organization of repeat reposting, and the growth of cascades that receive it.

\subsection{Single Source and Distributed Sources: Mechanism $\times$ Organization of Sources}\label{sec:primitive}
\begin{figure}[tb]
  \centering
  \includegraphics[width=\linewidth]{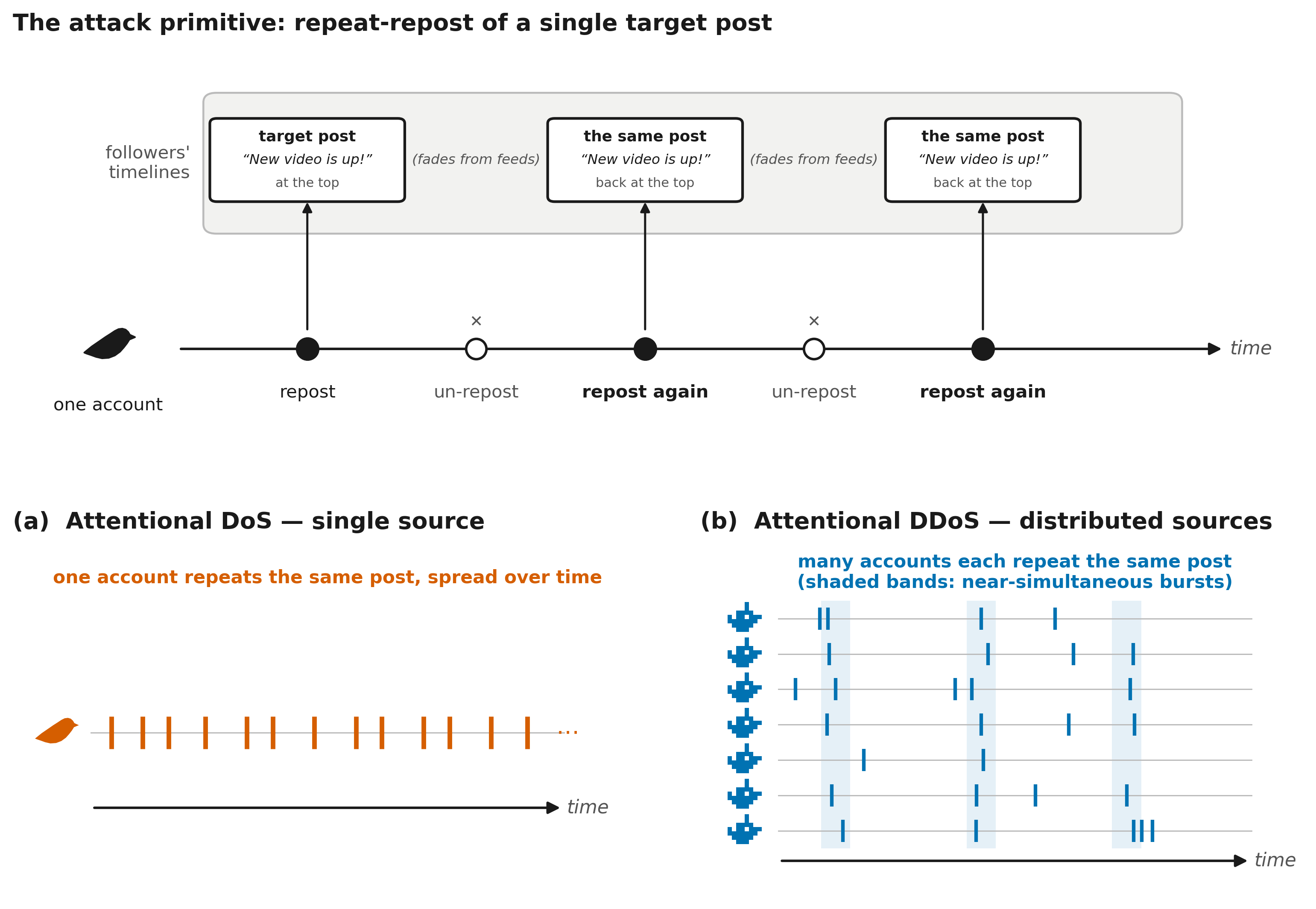}
  \caption{Top: the mechanism of repeat reposting. By repeating the cycle of repost, undo, and repost again, the same user can put the same post back into the followers' timelines any number of times, each time as a new share (display position and exposure are not observed). Bottom: raster view (one vertical bar = one repeat). (a) Single-source Attentional DoS: repeats by one account. (b) Distributed form: converging repeats by several habitual accounts; the time scale of synchrony is a continuous quantity.}
  \label{fig:schematic}
\end{figure}

Repeat reposting can be driven by one source or by many (Figure~\ref{fig:schematic}, bottom). The organization of sources is what the literature on coordinated amplification has measured, but that literature takes the mechanism of amplification (which operation is repeated) as given, and has not singled out repeat reposting as a distinct amplification mechanism (\nameref{sec:rel-coord}). DoS, on the other hand, describes the mechanism but says nothing about the organization of sources. This paper focuses on the crossing of the two axes: the mechanism (repeat reposting) $\times$ the organization of sources.

In security, the step from DoS to DDoS was ``the same mechanism, with distributed sources''. The same step can happen to repeat reposting, but many sources alone do not make organization: just as a flash crowd of normal users is not called DDoS, a large number of reshares of a company campaign is not an organized attack. We treat this by measuring, for every repeat repost, seriality (how habitually the same groups of accounts repeat reposts across many posts) and synchrony (how closely their actions are timed together), and we call the distributed form that separates out in the analysis Attentional DDoS (\nameref{sec:forms}). Below we write ADoS for Attentional DoS and ADDoS for Attentional DDoS (the full names remain inside the figure images).

\subsection{Research Questions}\label{sec:rq}
This is an exploratory study. No earlier work treats repeat reposting as a strategy of its own, so there was no hypothesis to state in advance about its distribution, its association with cascade growth, or its organization. We defined the questions and the design of the analysis step by step while looking at the full twelve months of data. The three questions below and the design of RQ2 (the definition of dose, the choice of controls, and the stratification by the repeater's reach and by the source structure) were fixed during that process, not before the results were seen. The findings of this paper are therefore presented not as the results of confirmatory tests but as hypotheses for confirmatory studies. RQ2 is at the center of the paper's claims.
\begin{itemize}
  \item RQ1 (distribution and scale): How common are repeat reposts, and how are they distributed across posts and accounts?
  \item RQ2 (the number of repeats and subsequent growth): How is the number of repeats early in the growth of a cascade related to the later arrival of new reactors and to the final size of the cascade? How does this relation change with the reach of the repeating account and with the source structure of the repeats (one account, several accounts, and if several, whether one is dominant)?
  \item RQ3 (organization and content): How is repeated amplification organized across the number of sources, in how habitually the same groups of accounts repeat reposts across many posts (seriality) and in how closely their actions are timed together (synchrony), and what does the content of amplified posts show about what is actually amplified?
\end{itemize}

% ---------- 02-related ----------
\section{Related Work}\label{sec:related}

\subsection{Scarce Collective Attention and the Measurement of Diffusion}\label{sec:rel-attention}
A model has shown that the fight for limited attention, together with the structure of the social network, makes the spread of information very uneven \cite{weng2012}. On the receiving side, large-scale data show that an overloaded timeline changes what a user passes on \cite{gomezrodriguez2014}. The final size of a cascade cannot be predicted from its early popularity alone; what matters is how the early adopters spread across communities \cite{weng2014}. The extra posts produced by the added exposure of trending are statistically significant but modest when they are estimated causally \cite{schlessinger2023}---the question that comes before our RQ2: whether additional visibility is associated with subsequent audience growth.

\subsection{Abuse of Reversible Operations and the Inflation of Engagement}\label{sec:rel-reversible}
Repeating a reversible operation has been used to inflate engagement metrics. An analysis covering more than a billion deletions on Twitter reports accounts that evade the daily posting limit by deleting posts, and coordinated networks that repeat likes and unlikes on the same post \cite{torreslugo2022}. Ephemeral astroturfing, in which bots post in coordination for a short time and delete immediately afterwards to fake a trend, involves a related use of reversible operations \cite{elmas2021}. Both are the same in form as this paper, in that repeating a reversible engagement operation inflates the apparent amount of engagement; they differ in the operation, which is deletion or the removal of a like. To our knowledge, no study has measured the repetition of undoing and redoing a repost itself.

\subsection{Coordinated Amplification and Its Detection}\label{sec:rel-coord}
Astroturf detection \cite{ratkiewicz2011}, bot amplification of low-credibility content \cite{shao2018}, and coordinated network detection \cite{pacheco2021} have measured how many accounts organize to amplify the same object, through the multiplicity of sources, co-action, and repeated participation (see \cite{ferrara2016} for a survey of social bots). More recent work estimates coordination as a degree rather than as a binary decision \cite{nizzoli2021}, clusters accounts of known identity by activity pattern and finds a group that does nothing but retweet \cite{keller2017}, describes the properties and the lifetime of retweet bots identified by buying them on a black market \cite{elmas2022}, and separates coordinated campaigns from organic trends using synchronized posting followed by immediate deletion as ground truth \cite{gopalakrishnan2025}. All of these infer organization from the temporal and structural signature of behavior.

\subsection{Flooding, Information Operations, and the Problem of Intent}\label{sec:rel-flooding}\label{sec:intent}
Flooding censorship \cite{king2017} and the firehose of falsehood \cite{paul2016} have documented strategies for changing an information environment through flooding, distraction, and repeated amplification. Coordinated reply attacks, in which accounts reply together to the posts of influential individuals and news organizations \cite{pote2025}, are on the same side as repeat reposting, in that they concentrate engagement on an existing post. But a behavioral signature alone cannot separate commercial spam, promotion, political manipulation, and the repeats of an enthusiastic individual. The classification by behavior and the substantive reading of what is amplified must therefore be kept apart in the analysis, and this paper does so by checking the content of amplified posts in order to classify it (\nameref{sec:content}).

% ---------- 02-methods ----------
\section{Materials and Methods}\label{sec:methods}

\subsection{Data}\label{sec:data}
\begin{itemize}
  \item Period and collection: April 2025 to March 2026 (twelve months), in Japan Standard Time (JST); all times and hours of day in this paper are JST. Posts and repost logs from cascades originating from posts classified as Japanese on X.
  \item Source: The data were collected via the X API. 
  The users and actions in the data are limited to those inside cascades that received at least 1,000 reactions (reposts, replies, quotes, and mentions). Data in smaller cascades are therefore not observed.
  \item Definition of ``Japanese-language'': the selection of the corpus depends on X's per-post machine language detection. The details of X’s per-post language detection method are not publicly documented, so its rules and accuracy cannot be independently verified. ``Japanese-language'' in this paper therefore means ``the posts that X's language detection treated as Japanese'', and we cannot say by what route posts written only in Chinese characters, without kana, entered the corpus. Our own analysis of language and script is done from the characters of the post text, independently of that detection.
  \item Scale: about 4.43 billion reposts (R) in the twelve months.
  \item Preprocessing: reposts with missing data were removed.
  \item Population for the effect analyses: in RQ2, we remove ADDoS targets whose root text is made only of Chinese characters without kana, or whose text could not be retrieved (56,943 targets, predominantly Chinese-script spam targets). They stay in the descriptions of distribution, organization, and content, where they are shown as a component.
  \item Ethics: this is an aggregate analysis of public data. We avoid re-identification and do not quote post text word for word; the four examples in Figure~\ref{fig:exemplars} carry only English summaries (to prevent re-identification by string search). The description of repeat reposting is an aggregate characterization of behavior that is already publicly observable.
\end{itemize}

\subsection{Definition of Repeat Reposting and Its Measures}\label{sec:def-repeat}
Among repost (R) actions, we group actions with the same user ID and referenced post ID into (user, target) pairs. For each pair, let $c$ denote the number of reposts. We count the second and subsequent reposts as repeat events. An account counts as a repeater of a target only when $c \geq 3$ (three or more reposts of the same post).  Because $c=2$ may reflect accidental repetition, we use $c \ge 3$ as a conservative threshold. The cascade-level measures are in Table~\ref{tab:metrics}.

\begin{table}[tb]
  \centering
  \caption{Measures of repeat reposting (per cascade)}
  \label{tab:metrics}
  \small\setlength{\tabcolsep}{3pt}
  \begin{tabular}{@{}p{0.40\columnwidth}p{0.53\columnwidth}@{}}
    \toprule
    Measure (symbol) & Definition \\
    \midrule
    Total repeats $R$ & number of reposts $-$ number of distinct (user, target) pairs \\
    Repeat ratio $\rho$ & $R$ / number of reposts = share of reposts that are repeats \\
    Max repeats per user $m_{\max}$ & largest number of repeats in one (user, target) pair \\
    Number of repeaters $n_{\mathrm{rep}}$ & number of repeaters ($c \geq 3$) \\
    Posts repeated $b$ (per account) & number of distinct targets the account repeated with $c \geq 3$ \\
    Serial amplifiers $k$ / their share $s$ & number / share of repeaters with $b \geq T = 100$ \\
    Synchrony $\sigma$ & share of accounts that acted in synchrony beyond what independent activity would give \\
    \bottomrule
  \end{tabular}
\end{table}

\subsection{Measures of Source Organization: Seriality and Synchrony}\label{sec:taxonomy}
The unit of analysis is the reposted post (target). Every target with at least one repeater is treated as an instance of Attentional DoS. The floor is the same as in the repeater definition ($c \geq 3$); we put no extra floor on the max repeats per user $m_{\max}$ or on the number of repeaters $n_{\mathrm{rep}}$, and describe them as continuous quantities. Two quantities measure the organization of sources; both are behavioral signatures that do not use the post text:

\begin{itemize}
  \item Seriality: per account, the number of posts repeated $b$ (the number of distinct targets repeated with $c \geq 3$ in the observation window); per target, the share of serial amplifiers $s$ (the share of repeaters with $b \geq T$, the serial amplifiers) and their number $k$. This is not evidence that the accounts acted together, but it captures repeated participation across multiple targets. The default $T$ is 100.
  \item Synchrony $\sigma$: how far the temporal co-occurrence of pairs of accounts (first repost times on the same target falling in a window $\Delta t$; we use $\Delta t = 60$ seconds, but this is only one instance of the minute scale and we give no meaning to the window length itself, since the daily scale is measured separately from repeat intervals) exceeds what independent activity would give. Pairs whose Poisson upper tail is significant under BH-FDR ($\alpha = 0.01$) with at least two co-occurrences are coordinated pairs, and for each target we take the share of accounts that belong to a coordinated pair and co-occurred. (We define it as excess over a null to avoid structural false positives from activity level, daily cycles, and arbitrary thresholds.) This is a continuous quantity, ``the strength of synchrony in the $\Delta t$ window'', not a binary one.
\end{itemize}

\subsection{Outcome Measures}\label{sec:outcomes}
All outcomes are computed from direct reactions to the target (of any type). Size: the reaction count (the total number of reaction events; an activity measure that includes repeats, not attention gained) and the number of unique reactors (the number of distinct accounts that reacted; an observable proxy for engaged reach, not views or impressions). We call the ratio reaction count / unique reactors the event multiplicity. It also exceeds 1 for legitimate multiple reactions other than repeat reposting, so multiplicity by itself is not inflation by repeats; we interpret systematically higher multiplicity in the repeated group as event-count inflation relative to unique-reactor counts. Speed: reactions per hour, time to peak, and the early-hour share (the share of reactions that came in the first hour).

Cascade growth is tracked by the cumulative number of unique reactors $K(t)$ from the start $t_0$ (the first reaction to the target). The outcomes for RQ2 are the arrival of new reactors from the end of the window to day 7, $K(t_0+7\mathrm{d}) - K(t_0+W)$; the 7-day size $K(t_0+7\mathrm{d})$ (about 0.1\% of cascades receive reactions after day 7, so this is in effect the final size); the share of cascades whose arrivals stop within 24 hours after the window; and the arrival rate by interval after the window (both unique reactors and reaction count).

\subsection{Analysis Procedure}\label{sec:method}
\begin{itemize}
  \item Operationalization of the forms (RQ1, RQ3): compute the measures per target and identify the distributed form (synchrony is kept as a continuous quantity).
  \item the number of repeats and subsequent growth (\nameref{sec:dose}): the unit is the cascade (target). The exposure is the number of repeat events $D_W$ and the number of repeaters $N_W$ in the exposure window $[t_0, t_0+W]$ from the start $t_0$ ($W \in \{1, 3, 6, 12, 24\}$ hours; 1, 3, and 6 hours are the main windows), and the source structure is defined by the share of the most active repeater $M_W/D_W$: single ($N_W = 1$), multi with one dominant ($M_W/D_W \geq 0.5$), and multi distributed ($< 0.5$). Treated cascades have a repeat in the window; controls are taken from a stratified sample of unrelated cascades without any repeat instance (331,090 over twelve months), as the nearest neighbor on the arrival curve, one to one, among cascades that agree on month, hour of start, weekday/weekend, size at the end of the window, and shape in the window (in \ref{sec:app-dose}). The estimator is the paired difference (logarithms for arrivals and size, shares for stopping), and intervals come from a bootstrap over cascade clusters.
  \item Reach of the repeating account (stratification for RQ2): follower counts are not observed, so we use prior arrival volume following the account's single reposts of other cascades as a proxy for reach. For an account, it is the median number of unique reactors that arrived within 6 hours after that account's single reposts of other cascades (cascades it did not repeat) in the 30 days before its repeat; it is missing when there are fewer than 5 such reposts. We compute it for the most active repeater in the window (the main repeater) at the time of its first repeat, and stratify by tertiles of its distribution on the treated side of the pairs. The proxy is available for about 90\% of the pairs. The dependence on reach and on dose is shown as elasticities, by regressing the paired log difference in 7-day size on $\log(1+P)$ and $\log(1+D_W)$ (paired bootstrap, 1,000 draws).
  \item Reach-matched controls (RQ2, supplementary): as controls we take cascades in which a repeater-type account (an account that made $c \geq 3$ repeats within the year) reposted exactly once in the window and did not repeat, add the quintile of that account's reach to the conditions above, and match treated cascades among those that agree on all conditions. The resulting paired difference compares repeated and single-repost cascades while holding the reach stratum fixed.
  \item Growth speed (RQ2): for the same pairs, we compare the arrival of unique reactors and the growth of the reaction count in five intervals after the window, $(W, W+1\mathrm{h}]$, $(W+1, W+3\mathrm{h}]$, $(W+3, W+6\mathrm{h}]$, $(W+6, W+24\mathrm{h}]$, and $(W+24\mathrm{h}, 7\mathrm{d}]$, by the paired difference in $\log(1+x)$. The reaction count includes the repeats themselves.
  \item Speed (RQ3): quantiles of time to peak and similar measures by form, and regressions of the difference between forms with unique reactors controlled.
  \item Time scale of synchrony (RQ3, \nameref{sec:spectrum}): besides minute-scale synchrony, we measure schedule-driven behavior from the regularity of inter-repeat intervals (IRI): pairs with $c \geq 5$ whose CV $= \sigma/\mu$ is below 0.5 and whose mean interval is about 24 hours are called a ``daily clock'' (a Poisson process gives CV $\approx 1$); we separate them from a daily habit by the spread of the hour of day, and count how many targets each clock account spans.
  \item Content check (RQ3, \nameref{sec:content}): after identification, the root text (first 200 characters) is checked against strict vocabularies for elections, politics, giveaways, money, and adult content, and for language by script. We obtained root text for 892,506 posts (99.71\% coverage; the 2,580 missing were all posted before the observation window) and also checked whether the poster's name follows a template (letters plus four or more digits). The composition of content is shown both by number of targets and weighted by repeat events.
\end{itemize}
% ---------- 03-results ----------
\section{Results}\label{sec:results}

\subsection{RQ1: Scale and Distribution of Repeat Reposting}\label{sec:rare}
Repeated reposts of the same post are rare: they appear in only 0.20\% of all (user, target) pairs, and they are typically spaced by hours to days (the median interval between consecutive repeats is about 9 hours; only about 6\% are under 60 seconds). This behavior is spread over 895,086 targets with a repeater ($c \geq 3$), 8,749,028 repeat pairs, and 357,044 repeating accounts. The distribution is skewed in two ways. First, on the target side: 652,914 targets (72.9\%) have exactly one repeater; the median number of repeaters $n_{\mathrm{rep}}$ is 1 (99th percentile 244, maximum 18,874), and the median max repeats per user $m_{\max}$ is 4 (99th percentile 32, maximum 4,437). Most repeats are light; the extreme ones sit in a long tail. Second, on the account side: only 8,685 accounts (2.43\%) are habitual accounts that repeated 100 or more posts ($b \geq 100$), but they hold 88.0\% of all repeat pairs.

The share of reposts that are repeats (the repeat ratio $\rho$) has mean 0.148, median 0.012, and 99th percentile 0.895; it falls monotonically from a peak near 0 (69.9\% below 0.1). Targets where repeats make up almost all reposts ($> 0.9$) number 7,723 (0.86\%); these targets are dominated by repeats from a single account (median number of unique reactors 1).

\subsection{RQ2: the number of repeats and subsequent growth}\label{sec:dose}
If repeat reposting keeps a post alive in the followers' timelines, cascades with more repeats in the early window should reach more new readers than cascades in the same state without repeats. The results do not support this expectation. The sign of the relation is decided not by the number of repeats but by who repeats, that is, by how many people usually react to reposts by the repeating account.

The population is 614,252 cascades with repeats (the ADoS population without Chinese-script spam ADDoS) and 234,850 cascades without repeats that satisfy the conditions on the start time. The share of cascades with a repeat in the window is 30\% for $W = 1$h, 47\% for 3h, 59\% for 6h, 73\% for 12h, and 87\% for 24h. Matching with equal conditions and nearest neighbors gives 9,992 pairs for $W = 1$h, 7,666 for 3h, 5,080 for 6h, 3,678 for 12h, and 2,452 for 24h (7.2\% down to 0.5\% of the treated side), with standardized differences of all matching variables at or below 0.05. The matching rate is low because most cascades with repeats have no cascade without repeats in the same state; the estimates apply to the treated cascades for which such a control exists. The reach proxy of the main repeater is available for about 90\% of these pairs; the tertile boundaries are 24 and 237 people (median 6-hour arrivals) for $W = 6$h, and 43 and 310 for $W = 1$h.

Table~\ref{tab:dose} shows the paired log difference in 7-day size by the main repeater's reach tertile and by source structure. Overall, the difference is negative for $W = 3$ to 12h ($-0.34$ to $-0.50$) and near zero for $W = 1$h and 24h. The negative side is concentrated in repeaters with low reach: for the low tertile, $-0.82$ to $-0.89$ for $W = 1$ to 6h (a little over 40\% of the control's size), and the share of cascades that stop within 24 hours after the window is 15 to 19 points higher. For the high tertile, the difference is small: $+0.26$ for $W = 1$h, zero for 3h, $-0.2$ for 6 to 12h, and $+0.06$ for 24h. Source structure acts independently of reach. Single-source cascades are negative ($-0.3$) for $W = 6$ to 12h even with high reach, while multi-source cascades (with one dominant repeater or distributed) are positive at high reach for every $W$ ($+0.1$ to $+0.2$) and negative at low reach, like single-source ones. Arrivals from the end of the window to day 7 show the same shape ($+0.16$ to $+0.85$ for high $\times$ multi distributed, $+0.25$ to $-0.49$ for high $\times$ single, $-0.98$ to $-1.28$ for low $\times$ single).

The dependence on the repeater's reach can be summarized in one elasticity. Regressing the paired log difference in 7-day size on the main repeater's reach $\log(1+P)$ and on the dose $\log(1+D_W)$ gives a reach elasticity of $+0.23$ [$+0.21$, $+0.25$] ($W = 1$h), $+0.18$ [$+0.16$, $+0.20$] (3h), and $+0.14$ [$+0.12$, $+0.17$] (6h) for unique reactors, and similarly positive values ($+0.20$, $+0.16$, and $+0.12$) for the reaction count. The dose elasticity is negative for unique reactors ($-0.23$, $-0.18$, $-0.13$); for the reaction count it is zero for $W \leq 6$h and positive only from 12h ($+0.13$ to $+0.14$, the repeats themselves being counted as reactions). That is, both the final number of readers and the reaction count of a repeated cascade grow in proportion to how many readers the repeater usually reaches, and not with the number of repeats. This dependence appears early. Within treated cascades only, with the size at the end of the window controlled, the reach coefficient is $+0.11$ for $W = 1$h, $+0.03$ for 3h, and zero from 6h on. The repeater's audience arrives in the first few hours, and reach adds nothing to the growth after that. The paired estimate stays positive in the 3 to 6 hour windows because the control cascade in the same state slows down afterwards, while the cascade of a high-reach repeater does not. This picture agrees with a coverage reading: the repeat delivers the same post again to the repeater's own audience, and the gain is bounded by the size of that audience. By source structure, the reach elasticity is positive for both single and multi ($+0.17$ to $+0.22$ at $W = 3$h), while the dose elasticity is negative for single and its interval includes zero for multi.

Reach-matched controls confirm this from the other side. When treated cascades are matched to cascades in which a repeater-type account of equal reach reposted once (2,364 down to 1,096 pairs), the difference in 7-day size stays near zero overall: $+0.18$ ($W = 1$h), $+0.02$ (3h), $-0.06$ (6h), $0.00$ (12h), and $+0.04$ (24h). It is positive at high reach ($+0.08$ to $+0.32$), negative at low reach ($-0.3$ for $W = 3$ to 6h), and positive for multi-source ($+0.1$ to $+0.26$). Once the reach of the repeating account is held equal, the contribution of repeating almost disappears; what remains is the positive association of repeats by high-reach accounts and by several accounts.

\begin{table*}[tb]
  \centering
  \caption{Early total repeats and the later 7-day size. Paired log difference [95\% interval] (number of pairs) between cascades with repeats and matched cascades without repeats, by the main repeater's reach tertile and by source structure.}
  \label{tab:dose}
  \footnotesize\setlength{\tabcolsep}{4pt}
  \begin{tabular}{@{}lrrr@{}}
    \toprule
    & $W = 1$h & $W = 3$h & $W = 6$h \\
    \midrule
    All & $-0.15$ [$-0.19$, $-0.12$] (9,045) & $-0.42$ [$-0.45$, $-0.38$] (6,880) & $-0.50$ [$-0.54$, $-0.45$] (4,557) \\
    \midrule
    Low reach & $-0.82$ [$-0.88$, $-0.75$] (3,021) & $-0.89$ [$-0.95$, $-0.82$] (2,295) & $-0.86$ [$-0.94$, $-0.79$] (1,522) \\
    Mid & $+0.10$ [$+0.04$, $+0.16$] (3,010) & $-0.32$ [$-0.38$, $-0.25$] (2,296) & $-0.39$ [$-0.47$, $-0.32$] (1,518) \\
    High reach & $+0.26$ [$+0.21$, $+0.32$] (3,014) & $-0.04$ [$-0.11$, $+0.02$] (2,289) & $-0.23$ [$-0.31$, $-0.16$] (1,517) \\
    \midrule
    High $\times$ single & $+0.27$ [$+0.21$, $+0.34$] (2,640) & $-0.07$ [$-0.15$, $-0.01$] (1,936) & $-0.33$ [$-0.41$, $-0.24$] (1,220) \\
    High $\times$ multi (one dominant) & $+0.20$ [$+0.07$, $+0.33$] (293) & $+0.13$ [$+0.01$, $+0.24$] (258) & $+0.19$ [$+0.07$, $+0.30$] (215) \\
    High $\times$ multi (distributed) & $+0.16$ [$-0.06$, $+0.37$] (81) & $+0.12$ [$-0.07$, $+0.29$] (95) & $+0.09$ [$-0.06$, $+0.24$] (82) \\
    Low $\times$ single & $-0.80$ [$-0.86$, $-0.73$] (2,870) & $-0.90$ [$-0.97$, $-0.83$] (2,094) & $-0.91$ [$-0.99$, $-0.82$] (1,371) \\
    Low $\times$ multi, one dominant & $-1.10$ [$-1.39$, $-0.83$] (133) & $-0.69$ [$-0.92$, $-0.46$] (176) & $-0.47$ [$-0.71$, $-0.25$] (135) \\
    \bottomrule
  \end{tabular}
\end{table*}

\begin{figure*}[tbp]
  \centering
  \includegraphics[width=0.92\linewidth]{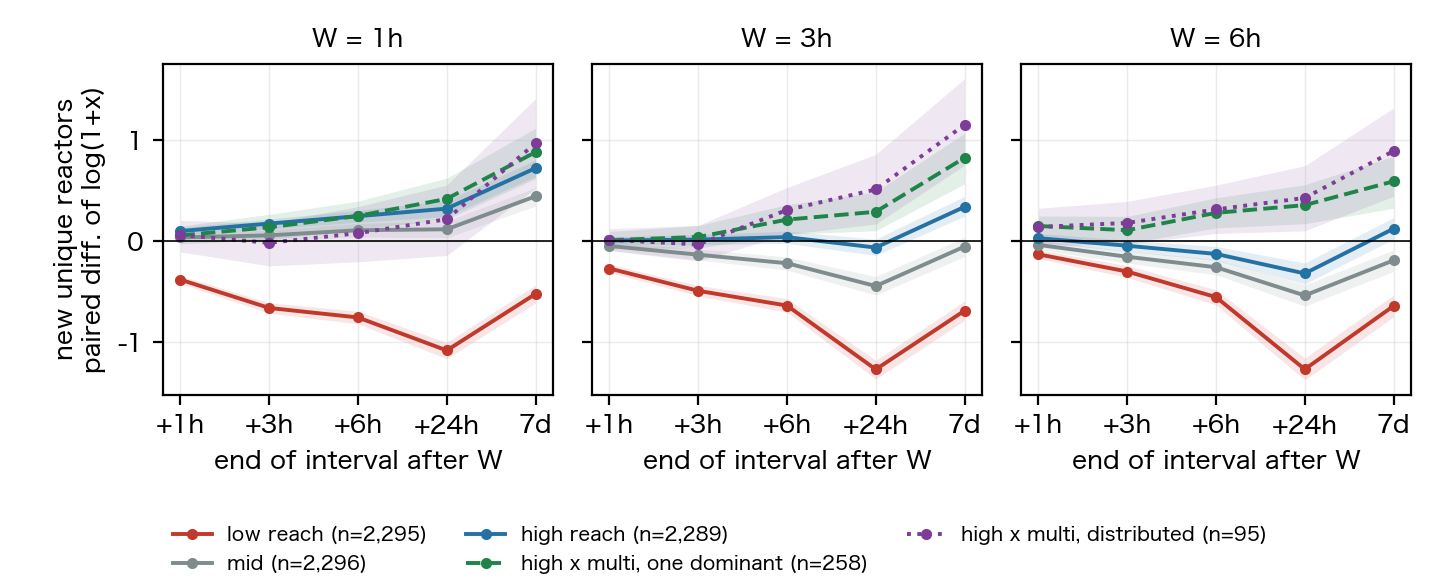}
  \caption{Growth speed after the window: paired difference in $\log(1+x)$ (treated $-$ control) of the arrival of unique reactors, for $W = 1$h, 3h, and 6h. Lines: the main repeater's reach tertile (low, mid, high) and high reach $\times$ multi-source (one dominant, distributed). The horizontal axis is the end of each interval after the window; bands are 95\% cascade-cluster bootstrap intervals. $W = 1$h is matched on only three points (10 minutes, 30 minutes, 1 hour), so it is coarser than the others, and its tertile boundaries are higher (43 and 310 people). }
  \label{fig:velocity}
\end{figure*}

Growth speed shows when the difference appears (Figure~\ref{fig:velocity}). Cascades of low-reach repeaters fall below the controls in the arrival rate of unique reactors right after the window; the gap deepens with time, is largest 6 to 24 hours after the window ($-1.27$, about a quarter of the control), and narrows after the first day as the cascades die out. At high reach there is no difference from the controls for 6 hours after the window, and only a shallow one afterwards. Multi-source cascades with high reach, by contrast, rise above the controls from 3 hours after the window, and the gap widens to $+0.6$ to $+1.15$ after 24 hours. In every stratum, the growth rate of the reaction count is $0.06$ to $0.21$ above that of unique reactors. This is because the repeats themselves are counted as reactions, so activity metrics always look better than the gain in readers.

These are descriptions, not causal effects. The choice to repeat and the effect of repeating cannot be separated, either by controls in the same state or by reach-matched controls (\nameref{sec:causal}). The whole picture can be explained by selection: low-reach accounts repeat cascades that are not growing, and several accounts repeat cascades that are. What can be said with confidence is this: a large number of repeats in the early window is not by itself linked to gaining new readers; where it is linked, the repeater already has high reach or several accounts are repeating; and even there, more repeats add nothing. Repeated amplification appears not as a proportional growth of readers but as continued activity around the same object. The inflation of activity metrics is a secondary consequence of this mechanism.

\subsection{RQ3: Organized Repeat Reposting}\label{sec:forms}
How much organization is there in this repeated amplification? We measure seriality and synchrony (\nameref{sec:taxonomy}) for the 895,086 repeat instances and look at how the distributed form appears.

Seriality is extremely concentrated. In contrast to the 72.9\% of targets with one repeater, 242,172 multi-source targets have more than one. The number of posts an account has repeated, $b$, follows a smooth power law with no natural break, but on the target side the distribution of the share of serial amplifiers separates into two regimes: in targets with three or more serial amplifiers, $s$ concentrates near 1, and in the near-continuous band of targets with 20 or more repeaters there is a valley at 0.37 to 0.44 (Appendix \ref{sec:app-sens}). Targets in which most repeaters are serial amplifiers can therefore be operationally distinguished from the rest. Synchrony, on the other hand, is continuous with no break (details in \nameref{sec:spectrum}). The feature that can be used to identify the distributed form is therefore seriality; synchrony stays a descriptive quantity.

We therefore call a target with three or more serial amplifiers and $s \geq 0.5$ an Attentional DDoS: several habitual accounts converge on the same post, a DDoS-like behavior. The minimum for ``distributed'' is three accounts rather than two because with a floor of two, most of the extracted set would be light targets in which two to four repeaters each repeat three or four times; the cut of 0.5 is a majority rule. The rest is split into single-source ADoS, with one repeater, and multi-source non-serial ADoS, with several repeaters but no seriality; the latter includes company campaigns (daily entries to giveaways) and normal users' reshares of a celebrity's request for shares. Table~\ref{tab:taxonomy} gives the counts.

\begin{table*}[tb]
  \centering
  \caption{Number of targets and share of repeat events by form (complete analysis, April 2025 to March 2026, $n = 895,086$)}
  \label{tab:taxonomy}
  \footnotesize\setlength{\tabcolsep}{4pt}
  \begin{tabular}{@{}lp{0.28\textwidth}rrrr@{}}
    \toprule
    Form & Rule & Targets & Share & \multicolumn{1}{c}{\shortstack[c]{Share of repeat events\\($c \geq 3$ pairs)}} & \multicolumn{1}{c@{}}{\shortstack[c]{Share of direct\\reactions}} \\
    \midrule
    Single-source ADoS & one repeater & 652,914 & 72.9\% & 7.8\% & 43.9\% \\
    Multi-source non-serial ADoS & $\geq 2$ repeaters, $< 3$ serial amplifiers or $s < 0.5$ & 157,368 & 17.6\% & 8.9\% & 40.3\% \\
    ADDoS & $\geq 3$ serial amplifiers and $s \geq 0.5$ & 84,804 & 9.5\% & 83.3\% & 15.8\% \\
    \bottomrule
  \end{tabular}
\end{table*}

\begin{figure*}[tbp]
  \centering
  \includegraphics[width=1.0\linewidth]{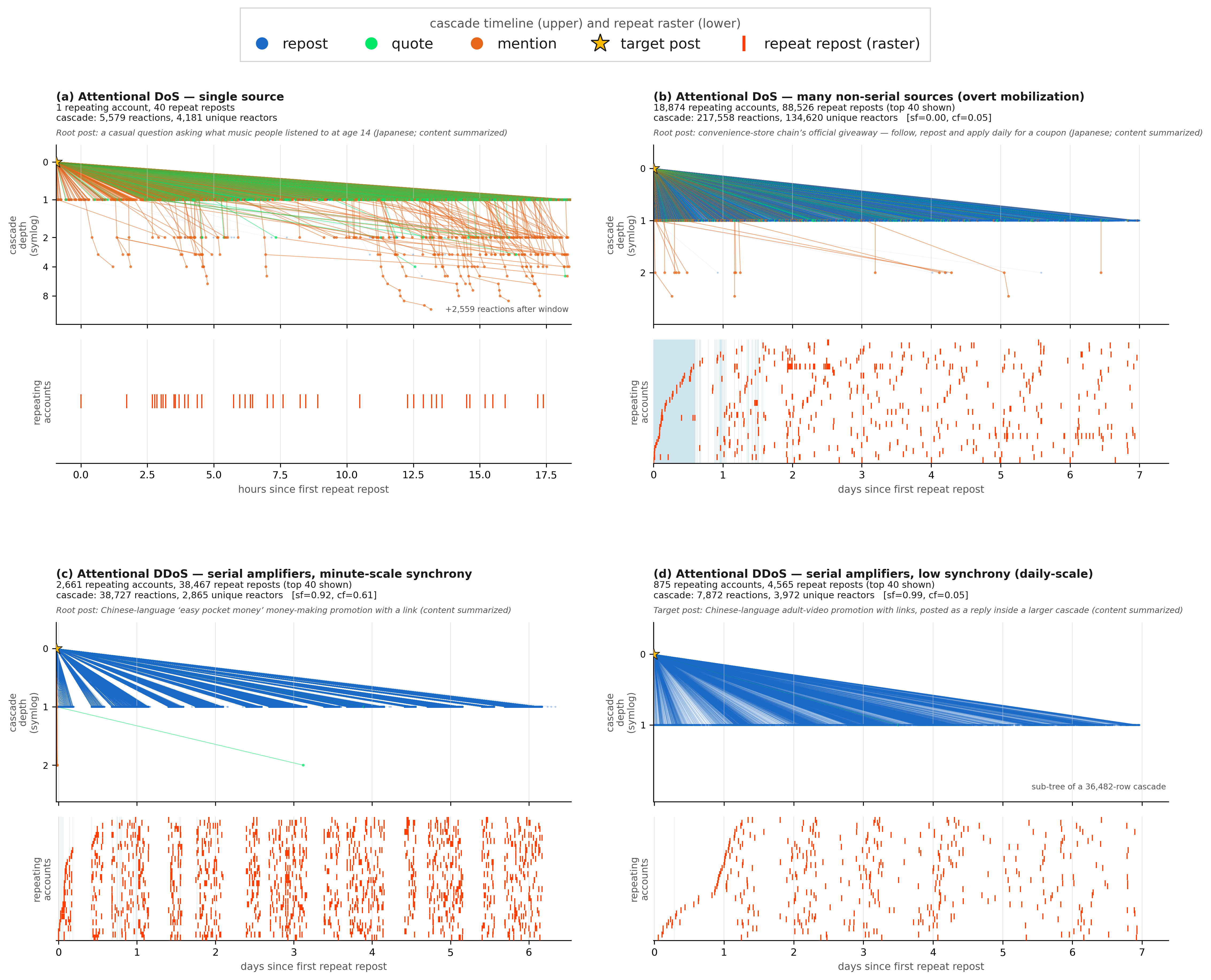}
  \caption{Examples by form (four panels; upper part of each: the cascade network, $\star$ = target, dots = reactions, lines = parent-child, colors = reaction type: repost blue, quote green, mention orange, vertical axis = depth (symlog); lower part: repeat raster, one red bar = one repeat, $c \geq 3$. The time axis starts at the first repeat and is shared. Root content is given only as an English summary.) (a) Single-source ADoS: an everyday question post repeated 40 times in 18 hours by one account; 5,579 reactions (reposts, quotes, and replies), 4,181 unique reactors, and an organic cascade more than 8 deep under the repeats. (b) Multi-source non-serial ADoS: a convenience-store chain's giveaway post (follow, repost, and enter daily) re-entered sparsely over 7 days by 18,874 accounts (the most in the period), with 134,620 unique reactors and an organic cascade of depth 2; seriality 0.00. (c) ADDoS, high synchrony [sf=0.92, cf=0.61]: a Chinese-language money-making post repeated in daily bursts (stripes) by 2,661 habitual accounts; a thin layer of depth 1. (d) ADDoS, low synchrony [sf=0.99, cf=0.05]: a Chinese-language adult-video promotion repeated daily without synchrony by 875 habitual accounts (the target is a reply inside a large cascade; only the subtree under it is drawn). Reading order (a) to (d): more sources alone do not make organization; seriality does, and that makes ADDoS.}
  \label{fig:exemplars}
\end{figure*}

The counts in Table~\ref{tab:taxonomy} are numbers of targets (repeated posts), not numbers of independent attackers or campaigns; as shown below, many targets are repeated by the same group of accounts. ADDoS is only 9.5\% of targets but holds 83.3\% of the $c \geq 3$ repeat events (28,717,953 of 34,495,057). Most repeats concentrate on the few posts on which groups of habitual accounts converge. This concentration rises steeply with the number of repeaters: the share of ADDoS is 14.3\% in the band of 2 to 4 repeaters, 70.6\% for 5 to 19, 85.7\% for 20 to 99, and 96.1\% for 100 or more. Single-source ADoS is not free of this serial infrastructure either: in 63.2\% of single-source targets, the only repeater is a serial amplifier. Examples are in Figure~\ref{fig:exemplars}.

The difference in organization also shows in the time signature: only ADDoS is a slow, persistent type (median time to peak 6 hours, against 0 hours for single-source and multi-source non-serial), and this slowness is not a by-product of size (Appendix~\ref{sec:app-figs}).

Inside ADDoS, seriality and synchrony are different signals, and synchrony spreads as a continuous quantity from 0 to above 0.9 (Figure~\ref{fig:spectrum}a). Multi-source non-serial ADoS has zero synchrony in 98.8\% of targets.

\subsection{The Time Scale of Synchrony and Its Shift}\label{sec:spectrum}\label{sec:clock}
Among the 84,804 ADDoS targets, synchrony $\sigma$ is 0 for 55.5\%, and the positive part spreads continuously from 0 to above 0.9 with no break (Appendix Figure~\ref{fig:spectrum}a). Minute-scale-dominated targets number 6,360 at the reference cut of 0.5 and 12,861 at 0.3; the count moves by a factor of two with the cut. This is the reason for removing synchrony from the identification conditions of ADDoS and keeping it as a descriptive quantity.

Synchrony is not limited to the minute scale. The distribution of inter-repeat intervals has a sharp spike at 24 hours; 17.8\% of pairs with $c \geq 5$ are clock-like, with CV below 0.5, and about two thirds of these have a mean interval of about 24 hours (Appendix Figure~\ref{fig:clock}). The hour of the repeats is not fixed (68.2\% of clock pairs drift), so a fixed-time daily habit alone is unlikely to explain this pattern; of the 5,329 targets with 10 or more clock repeaters, 4,823 are ADDoS; and 94\% of clock events fall on the targets of Chinese-script spam ADDoS. The daily-clock pattern is consistent with a spam amplification infrastructure operating on a shared or similar schedule.

This continuum also changes over the observation period. The share of minute-scale-dominated targets among new ADDoS falls from 19.2\% in April 2025 to 3.1 to 4.8\% from October 2025 on, and the median inter-repeat interval of pairs inside ADDoS moves from 18.9 hours toward a 24-hour clock (Appendix Figure~\ref{fig:spectrum}b). The account sets at the two ends overlap heavily (83.9\% of the accounts repeating at the minute-scale end also repeat at the clock end). This is consistent with the possibility that the same infrastructure shifted toward a coarser time scale, but without cohort tracking this interpretation remains a hypothesis.

What, then, does this moving infrastructure amplify, and what do single-source repeats amplify?

\subsection{Commercial Spam and Political Repeats}\label{sec:content}
All the classifications above are behavioral signatures that do not use the text. Since coordination does not identify intent, what content is associated with the identified forms is an independent question, and its answer decides how the classification is to be read. We checked the content of amplified posts (root text, 99.71\% coverage). The pattern differs depending on the denominator.

Weighted by repeat events, ADDoS is dominated by Chinese-script content that our vocabulary matching and sample inspection indicate is predominantly commercial spam.  Of the 28,601,451 ADDoS repeat events directed at targets with retrievable text (99.6\% of all), 98.2\% are directed at text made only of Chinese characters without kana, and only 0.42\% match political or election vocabulary. The matched types are money-making solicitations promising extreme returns (8.7\%) and links to adult content (10.6\%), in a pattern in which groups of accounts with template-like serial user names repeat each other's posts. The remaining four fifths of the events match none of the vocabularies; by sample inspection they are of the same kind (Chinese-language promotional templates whose wording our Japanese-leaning lists do not catch), and ``commercial spam'' in this paper refers to this whole set. A wider scan of the daily-clock side gives the same overall picture (37 election-related targets).

By number of targets, the picture is not uniform (Table~\ref{tab:content}). Of the 84,479 ADDoS targets, 33.0\% are Japanese with kana, and 6,589 (7.8\%) match political or election vocabulary. This component concentrates in light ADDoS with 3 to 19 serial amplifiers (political match rate 17.7\% in the band of 2 to 4 repeaters, 0.01\% for 100 or more); the typical case has a median of 4 repeaters, at most 6 repeats, and zero synchrony, that is, a few habitual accounts repeating a Japanese political post a few times each. The serial infrastructure that dominates the repeat surface by volume is commercial spam, but Japanese political repeats by small groups of serial amplifiers do exist inside ADDoS.

Politics is not absent from the repeat surface; most repeat reposts of political content come from a single account (the single-source form). Of the 151,409 targets that match political or election vocabulary, 80.9\% (122,468) are single-source ADoS, 14.8\% multi-source non-serial ADoS, and 4.4\% ADDoS. Of single-source ADoS, 18.8\% is political or election-related; the typical case is a single account repeating the same post, in support of or against a party or as a call to vote, 3 to 9 times (89.5\% of political targets have at most 9 repeats per user). The content of multi-source non-serial ADoS is, like single-source, mainly Japanese (92.2\%), and its strongest feature is giveaways at 19.8\%: normal users' reshares of company campaigns and of celebrities' requests for shares appear as multi-source repeats without seriality, however many sources they have. The content side, too, supports the picture that the number of sources alone does not make organization.

\begin{table*}[tb]
  \centering
  \caption{Content of repeated targets $\times$ form (full check, 99.71\% coverage). Vocabulary matching on the first 200 characters (exclusive priority: election $>$ politics $>$ giveaway $>$ money $>$ adult $>$ promotion $>$ news). ``Template name'' = letters plus four or more digits. The 2,580 missing texts were all posted before the observation window. The denominator of the repeat-event-weighted row is the repeat events on targets with retrievable text (99.6\% of all). The commercial-spam character of ADDoS targets that match no vocabulary is based on sample inspection (band of 100 or more repeaters).}
  \label{tab:content}
  \footnotesize\setlength{\tabcolsep}{3pt}
  \begin{tabular}{@{}lrrrrrrrr@{}}
    \toprule
    Form & $n$ (with text) & Japanese & Chinese-only & Political & Giveaway & Adult & Money & Template name \\
    \midrule
    Single-source ADoS & 650,980 & 95.1\% & 4.6\% & 18.8\% & 7.2\% & 1.2\% & 0.5\% & 14.8\% \\
    Multi-source non-serial ADoS & 157,047 & 92.2\% & 7.3\% & 14.2\% & 19.8\% & 1.6\% & 0.4\% & 11.1\% \\
    ADDoS (by targets) & 84,479 & 33.0\% & 67.0\% & 7.8\% & 11.7\% & 10.2\% & 0.8\% & 37.6\% \\
    ADDoS (by repeat events) & 28,601,451 & 1.8\% & 98.2\% & 0.4\% & 0.6\% & 10.6\% & 8.7\% & — \\
    \bottomrule
  \end{tabular}
\end{table*}

Limitations. The matching is vocabulary-based and may undercount political and election-related content. The commercial-spam character of ADDoS targets that match no vocabulary is a judgment from sample inspection; systematic manual coding of a random sample is future work. The most important reservation is this: what this section shows is that, on the repeat surface of this data, the serial infrastructure that dominates by volume shows no political intervention, and repeat reposts of political content appear in single-source and light serial forms. It does not show the absence of political coordinated inauthentic behavior (\nameref{sec:who}).

% ---------- 04-discussion ----------
\section{Discussion}\label{sec:discussion}

\subsection{Amplification Is Not Diffusion}\label{sec:trap1}
What the results show is the pattern associated with a behavioral mechanism, repeat reposting. Contrary to the view that cascades with more repeats in the early window should reach more new readers than cascades in the same state without repeats, the sign of the observed relation was decided not by the number of repeats but by who repeated, that is, by how many people usually react to reposts by the repeating account. Cascades repeated by low-reach accounts subsequently grew more slowly than matched controls, whereas those repeated by high-reach accounts or by several accounts showed more sustained growth; with reach held approximately equal, the difference between repeated and single-repost cascades was near zero. The observed amplification agrees not with a widening of the audience but with putting the same object back into the competition for visibility again and again, a repeated demand on a finite shared resource, collective attention (\nameref{sec:bg}). Observed amplification therefore cannot be automatically read as diffusion. A secondary consequence is that raw event counts overstate the number of people involved (the growth rate of the reaction count was above that of unique reactors in every stratum), so before reading a reaction count or repost count as attention or reach, one must check whether the event sequence contains repeat reposts.

That the final reach and reaction count of a repeated cascade grow with the repeater's usual reach and not with the number of repeats agrees with the reading that the gain from repeating comes from delivering the post again to the repeater's own audience (coverage). In this data, however, we do not observe whether the added readers are the repeater's followers. The observed implications of Attentional DoS should be described not as the taking of reach from other content, but as the repeated coverage of a bounded audience and as the self-reinforcement of posts that are already growing. The self-reinforcement side (the positive association when several high-reach accounts pile repeats on a growing post) carries the reservation that it cannot be separated from selection.

\subsection{Implications for Coordination Detection}\label{sec:coordrec}\label{sec:who}
ADDoS is supported by groups of habitual accounts, but temporal synchrony is continuous, its time scale spreads from minutes to a day, and it is shifting toward the coarse side (\nameref{sec:spectrum}). The count of ``coordination'' is therefore a function of the definition of the estimator (6,360 to 12,861 with the minute-scale threshold, 84,804 for ADDoS identified by seriality), and a count reported without its time scale and threshold can be difficult to interpret. Three suggestions for coordination detection: (a) Add a mechanism-specific feature, ``convergence on repeat reposting''. (b) Separate seriality from synchrony; use seriality for identification and measure synchrony as a continuous quantity against a null of independent activity. (c) Measure synchrony at several time scales and report it with the time scale; one fixed window may miss infrastructures whose characteristic time scale changes over time.

In this data, the organized repeat amplification that dominates the repeat surface by volume is primarily associated with commercial spam rather than political content (\nameref{sec:content}). A coordination detector may therefore capture large volumes of commercial spam alongside, or instead of, political information operations. This is not proof of the absence of political coordination: political repeats exist at large scale (over 150,000 targets), most in the single-source form and some as light ADDoS by a few serial amplifiers, and both are difficult to detect using co-occurrence-based methods. Coordination detection needs to be designed across several attack primitives as well as across time scales.

\subsection{Implications for Platform Design}\label{sec:platform}
Where the mitigation approach sits depends on the number of sources. In single-source ADoS the repeated activity sits in one account, so mitigation can be done per source account: demotion of display weight for repeats of the same (user, post) pair (without hurting the legitimate first share), rate limits, and making repeats visible. In ADDoS the number of repeats per account may be too small for purely per-account thresholds to capture the distributed pattern, so the relations between accounts must be used (identifying groups of serial repeaters that converge on the same target). For the same phenomenon, single sources call for within-node measures and ADDoS for between-node measures. The concentration of most of the repeat surface in a small group of habitual accounts also means that the target of between-node measures is small. All of these are interventions that reduce, or make visible, repeated demands on visibility by the same content. On top of them come metric design and transparency: treat and present activity metrics (event counts) separately from unique reactors, distinguish repeated-event counts from unique-reactor counts in user-facing metrics, and design coordination monitoring at several time scales so that it survives the shift (\nameref{sec:coordrec} (c)).

\subsection{Limitations}\label{sec:limitations}

\paragraph{Direction of causality and the limits of observational data.}\label{sec:causal}
The direction of causality between repeats and growth is not identified. The matched comparison of RQ2 is observational and cannot separate the effect of repeating from selection: ``low-reach accounts push cascades that are not growing'' and ``several accounts push cascades that are''. The matching rate is 0.5 to 7\% of the treated side. The reach proxy is the volume of arrivals after a reaction, not a follower count, and the main repeaters without a proxy (about a tenth) fall out of the stratification. Controls are drawn from a stratified sample of unrelated cascades, so the control pool is constructed differently from the treated population; the matching procedure is intended to reduce these observed differences. The data are limited to cascades with at least 1,000 reactions (\nameref{sec:data}), so the consequences of repeats in small cascades are not observed. We also do not observe impressions, timeline positions, or actual views. The study is exploratory (\nameref{sec:rq}), and chance fits from the choice of windows, thresholds, and strata cannot be excluded. 

\paragraph{Behavioral signatures, not verdicts on intent or authenticity.}\label{sec:signature}
Our classification identifies the existence of (user, post) repeats and the behavioral signatures of the number of repeaters and seriality; it does not decide whether an account is a bot or what its intent is. For the political repeats of single-source ADoS, these data (without followers or account age) cannot separate organized manipulation from an enthusiastic individual.

\paragraph{Thresholds and generalization.}\label{sec:sens}
The sensitivity to the thresholds of this paper (repeater floor $c \geq 3$, serial-amplifier floor $T = 100$, minimum of 3 serial amplifiers for ADDoS, seriality cut 0.5) is given in Appendix \ref{sec:app-sens}. Because $c \geq 3$ keeps light repeats in the population, ``extremeness'' must be read as a description of the tail, not a property of the population. The number of targets identified by synchrony $\sigma$ varies substantially with the choice of threshold, so we report this dependence itself as a result and keep synchrony out of the identification conditions. The data are limited to one platform and one language sphere.

% ---------- 05-conclusion ----------
\subsection{Conclusion}\label{sec:conclusion}
Repeat reposting puts the same post back into the followers' timelines again and again without making a new post, thereby allowing repeated entry into the competition for visibility without producing new content. This paper named this practice Attentional DoS and covered 895,086 repeat-reposting instances in the observed corpus over twelve months, drawn from 4.43 billion reposts in cascades originating from posts classified as Japanese on X. Subsequent audience growth is associated less with the number of repeats than with the estimated reach of the repeating accounts. Repeat reposting is dominated by a small group of habitual accounts; the distributed form identified by seriality (ADDoS) holds 83\% of repeat events, and its synchrony is a shifting continuum from minutes to a day. Most repeat reposts of political content come from a single account (the single-source form), which detection of coordination between accounts cannot see by construction. The observed pattern is better characterized not as the taking of reach from other content, but as repeated exposure opportunities within a bounded audience and the self-reinforcement of posts that are already growing. The analysis is exploratory, and the findings are hypotheses for confirmatory studies. Future work includes confirmation with a design fixed before the results are seen, causal identification through natural experiments, direct measurement of occupation with display and exposure data, bot attribution from account features, and generalization to other languages and platforms.

\section*{Acknowledgments}
\ACKTEXT

% aaai2026.sty sets the bibliography style itself when natbib is loaded (do not add \bibliographystyle)
\bibliography{references}

\CHECKLIST

\appendix
\setcounter{secnumdepth}{1}
% ---------- 07-appendix ----------
% Appendix (outside the page limit at ICWSM 2027)

\section{Thresholds and Sensitivity}\label{sec:app-sens}

\begin{itemize}
  \item Repeater floor $c$: with $c \geq 2$ there are 2,899,085 ADoS instances and 452,974 ADDoS (15.6\%; this includes mistakes, and because $b$ is also counted with $c \geq 2$ the number of habitual accounts rises to 20,583). With $c \geq 3$: 895,086 and 84,804 (9.5\%); $c \geq 4$: 462,634 and 40,259 (8.7\%); $c \geq 5$: 295,768 and 25,388 (8.6\%). Raising the floor does not move the share of ADDoS, and the reason for $c \geq 3$ (excluding mistakes) stays.
  \item Floor $T$ on posts repeated ($k = 3$, cut 0.5): with $T = 20 / 50 / 100 / 200 / 500$, ADDoS numbers 101,807 / 92,613 / 84,804 / 74,864 / 62,490. The distribution of $b$ over accounts is power-law-like (exponent about 1.55, 95th percentile 19, 99th percentile 574, maximum 14,600) with no natural break, so $T$ is a conventional value.
  \item Minimum number of serial amplifiers $k$ ($T = 100$, cut 0.5): with $k = 2 / 3 / 5 / 10$: 148,158 / 84,804 / 55,615 / 41,965. With $k = 2$, 59\% of the extracted set are light targets with 2 to 4 repeaters. Multi-source non-serial ADoS contains 63,354 targets with exactly two serial amplifiers, which is the difference between $k = 2$ and $k = 3$.
  \item Seriality cut ($T = 100$, $k = 3$): with 0.3 / 0.5 / 0.7 / 0.85: 86,939 / 84,804 / 78,821 / 69,468. The valley near 0.85 in the overall histogram of $s$ is an artifact of discreteness in targets with few repeaters, where the value sticks to 1/2, 2/3, and 5/6. In the near-continuous band of 20 or more repeaters the valley is at 0.37 to 0.44, and the cut of 0.5 was placed just above it as a majority rule. Over the 144 grid points of $T$, $k$, and the cut, the counts are monotone without exception.
\end{itemize}

\section{Details of the Matching}\label{sec:app-dose}

In both groups the start must be at least 24 hours after the start of the month and 7 days before its end, and $K_W$ must be at least 5. Among the cascades that agree on the cell variables of \nameref{sec:method} (the size and the shape in the window entering as decile and quintile), we take the one with the smallest Mahalanobis distance on the cumulative arrival curve in the window (the logarithm at each point from 10 minutes to $W$) and on the time since the last arrival, with a caliper of 0.25 SD, one to one and without replacement. The standardized differences of all matching variables are at or below 0.05, and intervals are Poisson bootstrap over cascade clusters (1,000 draws).

\section{Time Signature}\label{sec:app-figs}

This slowness is not a by-product of size: regressing $\log_{10}(1+h)$ of the time to peak (in hours) with unique reactors controlled (HC1, $R^2 = 0.084$) gives a small coefficient for unique reactors ($+0.047$) and form coefficients of $+0.45$ for ADDoS (about 2.8 times slower than single-source) and $-0.07$ for multi-source non-serial.

\begin{figure*}[tbp]
  \centering
  \includegraphics[width=0.92\linewidth]{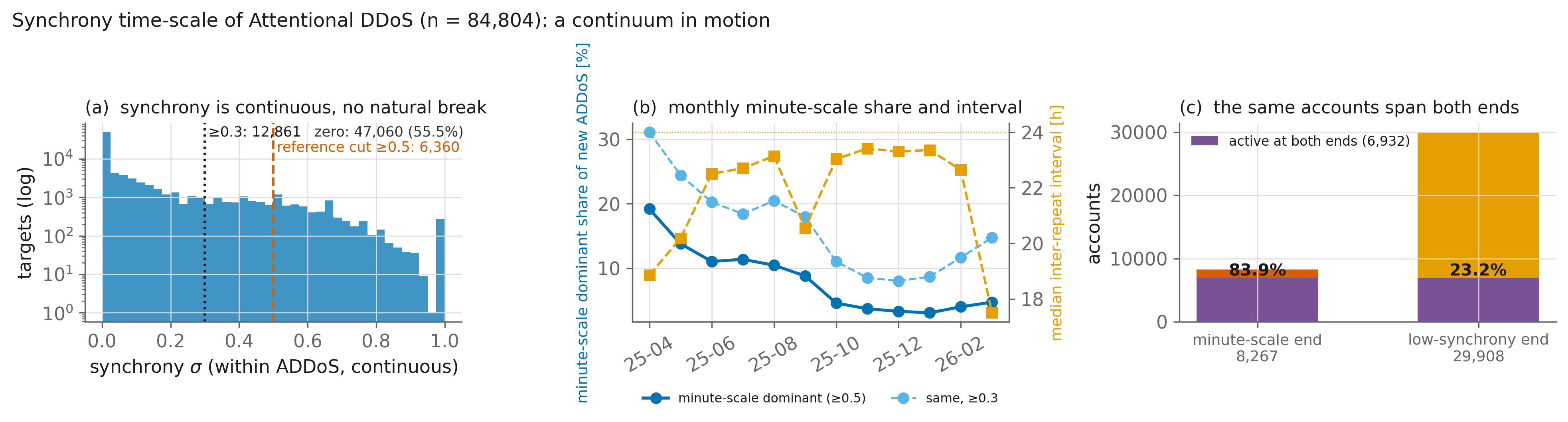}
  \caption{The time scale of ADDoS synchrony is a moving continuum ($n = 84{,}804$). (a) Distribution of synchrony $\sigma$ (log vertical axis): more than half are zero, and the positive part is continuous with no break; every cut is a cut through a continuum. (b) The share of minute-scale dominated targets (blue) collapses from 19.2\% to 3 to 5\%, while the median inter-repeat interval of pairs (orange) approaches 24 hours. (c) 83.9\% of the accounts repeating at the minute-scale end also repeat at the clock end.}
  \label{fig:spectrum}
\end{figure*}

\begin{figure}[tb]
  \centering
  \includegraphics[width=\linewidth]{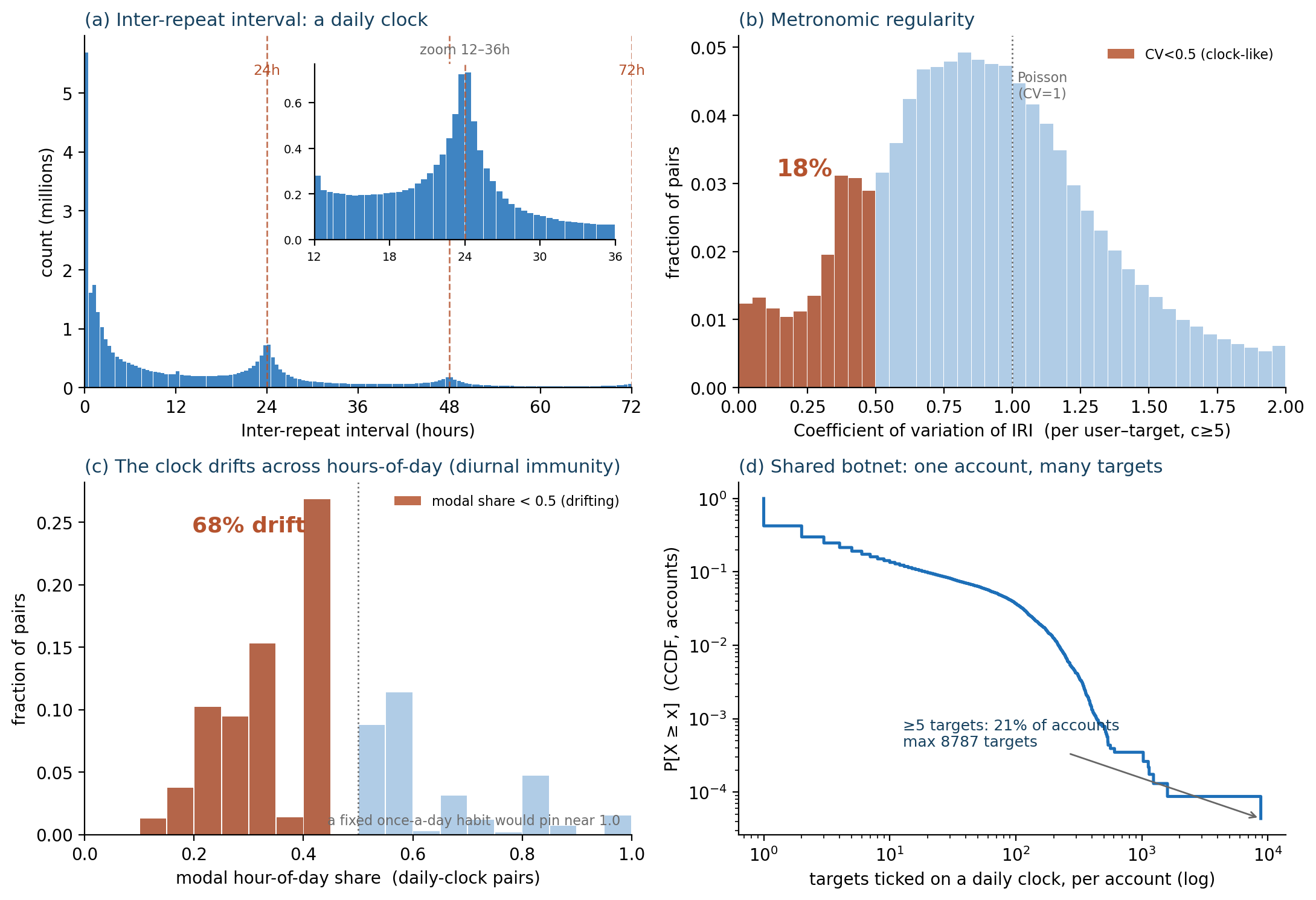}
  \caption{Structure of inter-repeat intervals (whole period, $c \geq 3$, \nameref{sec:spectrum}). (a) IRI histogram: spikes at 24, 48, and 72 hours. (b) Distribution of CV: 18\% are clockwork ($< 0.5$). (c) Share of the most frequent hour for clock pairs: 68\% drift and cannot be explained by a fixed-time habit. (d) Targets spanned by clock accounts: up to 8,787 (a shared schedule group).}
  \label{fig:clock}
\end{figure}

\end{document}